\documentclass[
]{ceurart}

\usepackage{listings}
\usepackage{svg}
\usepackage{adjustbox}
\begin{document}

\copyrightyear{2024}
\copyrightclause{Copyright for this paper by its authors.
  Use permitted under Creative Commons License Attribution 4.0
  International (CC BY 4.0).}

\conference{Accepted in the 14th International Conference on Learning Analytics and Knowledge (LAK24)}

\title{Experimental Interface for Multimodal and Large Language Model Based Explanations of Educational Recommender Systems}


\author[1]{Hasan Abu-Rasheed}[%
email=hasan.abu.rasheed@uni-siegen.de
]
\cormark[1]
\address[1]{University of Siegen, Siegen, Germany}

\author[1]{Christian Weber}[
]
\author[1]{Madjid Fathi}[
]

\cortext[1]{Corresponding author.}

\begin{abstract}
  In the age of artificial intelligence (AI), providing learners with suitable and sufficient explanations of AI-based recommendation algorithm’s output becomes essential to enable them to make an informed decision about it. However, the rapid development of AI approaches for educational recommendations and their explainability is not accompanied by an equal level of evidence-based experimentation to evaluate the learning effect of those explanations. To address this issue, we propose an experimental web-based tool for evaluating multimodal and large language model (LLM) based explainability approaches. Our tool provides a comprehensive set of modular, interactive, and customizable explainability elements, which researchers and educators can utilize to study the role of individual and hybrid explainability methods. We design a two-stage evaluation of the proposed tool, with learners and with educators. Our preliminary results from the first stage show high acceptance of the tool’s components, user-friendliness, and an induced motivation to use the explanations for exploring more information about the recommendation.
\end{abstract}


\maketitle

\section{Introduction}

Explaining educational recommendations to learners showed promising potential to enhance the learner’s acceptance of the learning recommendations \cite{ooge_explaining_2022}. Explaining the recommendation is not only meant to clarify why certain content is recommended but also to support the learner’s ability to make an informed decision about following the recommendation. This ability is greatly influenced by the type and volume of information learners receive from the explanations. Multiple types of explanations have been investigated in recent years. Hybrid and multimodal explanation approaches were found to increase the learner’s satisfaction \cite{pecune_model_2019} and engagement with the system \cite{tsai_evaluating_2019}. With the new capabilities of LLMs, an emerging form of conversational explanations holds the potential to offer a better understanding of the recommendation by engaging the learner in a question-and-answer session about the recommended content. While the use of chatbots in education is not new, the use of LLM-powered chatbots in generating learning explanations is still under investigation, due to the great limitations of LLMs in a sensitive field like education. This reveals an essential and urgent need for creating tools that support researchers and educators in experimenting and evaluating different types of explanation modalities and approaches, as well as the different hybridization approaches amongst them. Our main contribution in this research is developing such an experimental tool, in the form of a modular and interactive interface, which allows the delivery of multimodal and chatbot-supported explanations.

\section{Experimental Interface of Recommendation Explainability}
We propose an interactive, experimental, web-based interface as a tool for delivering textual, visual, and chatbot-supported explanations of learning-recommendations. Our design is based on a pedagogical analysis of requirements expected from a learning-explanation. Pedagogy experts and educators were interviewed to determine these requirements, and how the tool’s design can be tailored to visualize them to the learners. As a result, we focus on providing the means to explain: 1) the recommendation’s connection to the learning goal, 2) relations amongst the recommended topics, 3) connection to the user-profile, and 4) connection to the teacher’s original structure of the materials. Additionally, open-ended questions are supported through a chatbot. Our components, see Figure 1, are designed to enable experimenting with different combinations of explainability methods, by selecting which components are visible to the learner. We offer the following explainability components for experimentation: 
\begin{itemize}
    \item \textbf{Textual explanations}: which are provided in a variable-length text area. By incorporating Markdown, this component allows for versatile text presentations.
    \item \textbf{Tag-based explanation}: this is a more specific form of textual explanation but allows the user to click on the tags to reach further information using hyperlinks.
    \item \textbf{Hierarchical structure}: which allows a clear overview of a recommendation’s hierarchy, such as the one created by a human educator. It offers expandable elements and clickable titles with hyperlinking.
    \item \textbf{Graph-based visual explanations}: it provides interactive graphs for visual explanations, which can also show textual elements on the graph.
    \item Radar charts: offering the potential to create multidimensional comparisons, to explain, e.g., coverage of the recommendations or overlap with preferences.
    \item \textbf{Venn diagrams}: which provide a more focused overlap view, with a limited number of dimensions, but with interaction features including clicking, hover-over, and numeric or textual overlays. 
    \item \textbf{Chatbot-based explainability}: This component offers flexible support for chatbot use in explainability. We support LLMs using OpenAI API. Our support of LLMs is also achieved by the design of chat-messages to include contextual information and human defined rules, offering more control over the LLM’s output, and thus reducing the risk of hallucinations and irrelevant responses. Our chatbot element enables researchers and educators to utilize the information on other elements of the interface, to enrich the LLM’s prompt and context. It also provides a connection to the database, to acquire additional information about the recommended items. A multi-agent chatting support is also offered, allowing the user to chat with other users, mentors, or even different LLMs. This provides high flexibility for experimenting and evaluating chatbot roles in recommendation explainability.
\end{itemize}

\begin{figure}
  \centering
  \includegraphics[width=\linewidth]{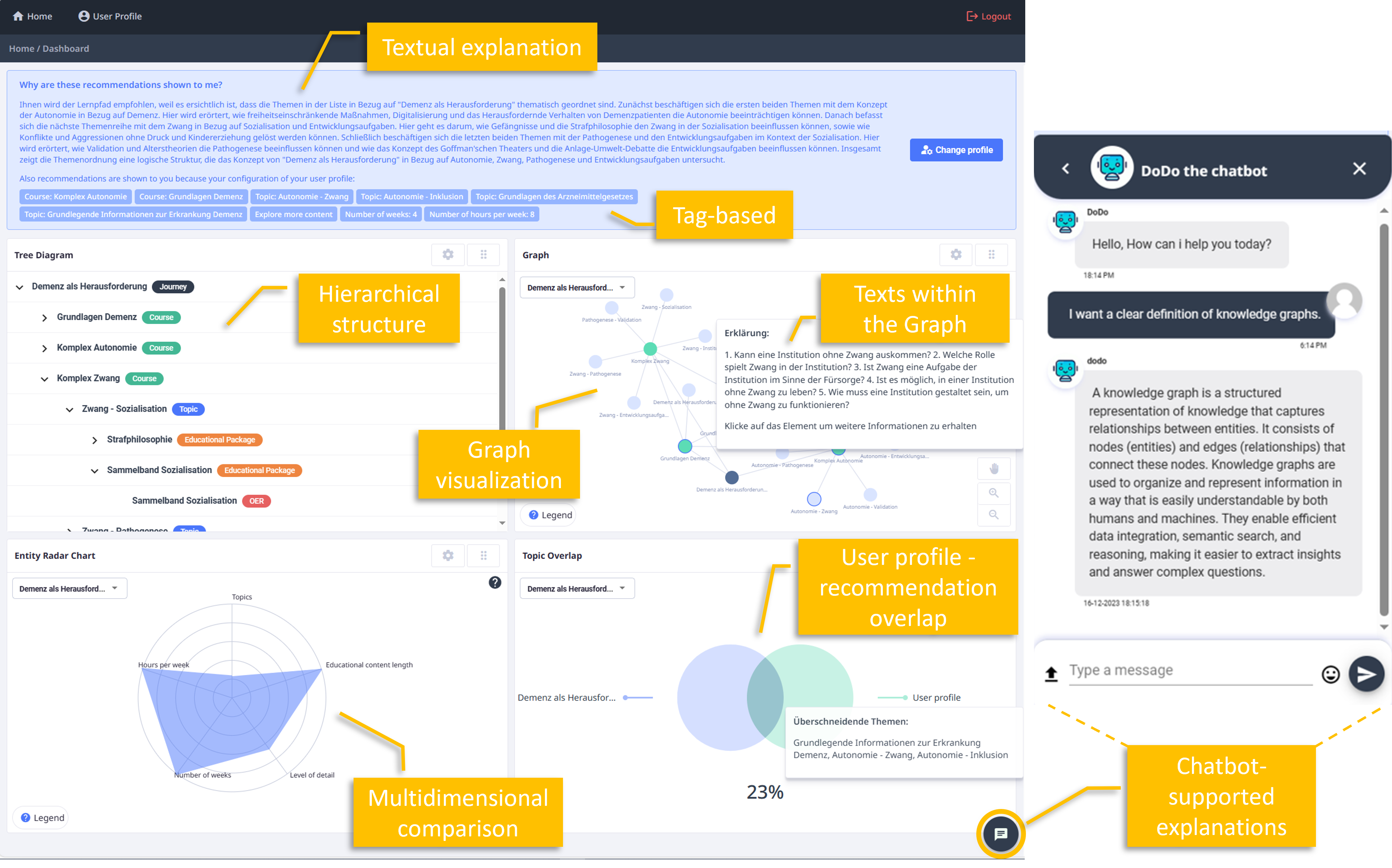}
  \caption{Proposed interface with the complete view of explainability components.}
\end{figure}

\section{Evaluation and Results}

We designed a two-stage evaluation strategy for the proposed tool. The first stage is to evaluate the tool’s usability by the learners. The second stage is to evaluate the tool’s usability by educators, and its features for accommodating singular and hybrid explainability approaches. We conducted the first evaluation stage with learners in the form of focus groups, in which learners were asked to perform a set of learning tasks on the interface. The learners then filled out a survey focusing on 1) their acceptance of the components, 2) the tool’s user-friendliness, 3) the ease of retrieving information from the components, and 4) the tool’s role in motivating the user to explore more about the recommendation. The second stage is planned in the next step of this ongoing research. In this stage, educators will test the tool with a set of pre-defined tasks. Qualitative interviews will be conducted to survey the educators’ feedback on 1) the tool’s usability, 2) most important features, 3) features that should be added to accommodate their experimentation requirements, and 4) most importantly, if the tools provide the sufficient level of abstraction for educators with a non-technical background to understand and utilize its features with minimal support from developers.

\section{Conclusion}
In this research, we propose an experimental tool that offers a comprehensive set of explainability components and features, to enable educators and researchers to test individual and hybrid explanation modalities and methods. Our tool is a web-based interface with interactive textual and visual modules. We introduce a chatbot element for conversational explanations, and offer thorough support for LLM-based chatbots, through accommodating OpenAI’s API, prompt contextualization, rules, and connection to the databases. As ongoing research, we are continuing our evaluation plan with educators, and extending the features of the tool to include new visualization modules and more support for other LLM APIs, such as LlaMa and Gemini.

\bibliography{sample-1col}

\end{document}